\documentclass[3p, twocolumn, authoryear]{elsarticle}
\usepackage{lmodern}
\usepackage{amssymb,amsmath}
\usepackage{ifxetex,ifluatex}
\usepackage[T1]{fontenc}
\usepackage[utf8]{inputenc}
\usepackage[unicode=true]{hyperref}
\usepackage{xcolor}
\usepackage{lineno}
\usepackage{setspace}

\journal{arXiv}

\begin{document}
\begin{frontmatter}

\title{The role of thickness inhomogeneities in hierarchical cortical folding}
\author[ICS,INM]{Lucas da Costa Campos}
\author[ICS]{Raphael Hornung}
\author[ICS]{Gerhard Gompper}
\author[ICS]{Jens Elgeti\corref{t1}}
\ead{j.elgeti@fz-juelich.de}
\author[INM,JARA,HHU]{Svenja Caspers\corref{t1}}
\ead{s.caspers@fz-juelich.de}

\cortext[t1]{The last authorship is shared between these authors.}

\address[ICS]{Theoretical Physics of Living Matter, Institute of Biological Information Processing (IBI-5), Research Centre Jülich, Jülich, Germany}
\address[INM]{Institute of Neuroscience and Medicine (INM-1), Research Centre Jülich, Jülich, Germany}
\address[JARA]{JARA-Brain, Jülich-Aachen Research Alliance, Jülich, Germany}
\address[HHU]{Institute for Anatomy I, Medical Faculty, Heinrich-Heine University, Düsseldorf, Germany}
\date{April 1, 2020}

\begin{abstract}
The morphology of the mammalian brain cortex is highly folded. For long it
has been known that specific patterns of folding are necessary for an optimally
functioning brain. On the extremes, lissencephaly, a lack of folds in humans, and
polymicrogyria, an overly folded brain, can lead to severe mental retardation,
short life expectancy, epileptic seizures, and tetraplegia. The construction of
a quantitative
model on how and why these folds appear during the development of the brain is
the first step in understanding the cause of these conditions.
In recent years, there have been various attempts to understand and model the
mechanisms of brain folding. Previous works have shown that mechanical
instabilities play a crucial role in the formation of brain folds, and that the
geometry of the fetal brain is one of the main factors in dictating the folding
characteristics. However, modeling higher-order folding, one of the main
characteristics of the highly gyrencephalic brain, has not been fully tackled.
The effects of thickness inhomogeneity in the gyrogenesis of the
mammalian brain are studied \emph{in silico}. Finite-element simulations of
rectangular slabs are performed. The slabs are divided into two distinct
regions, where the outer layer mimics the gray matter, and the inner layer the
underlying white matter. Differential growth is introduced by growing the top
layer tangentially, while keeping the underlying layer untouched. The brain
tissue is modeled as a neo-Hookean hyperelastic material. Simulations are performed with
both, homogeneous and inhomogeneous cortical thickness.
The homogeneous cortex is shown to fold into a single wavelength, as is
common for bilayered materials, while the inhomogeneous cortex folds into
more complex conformations. In the early stages of development of the
inhomogeneous cortex, structures reminiscent of the deep sulci in the
brain are obtained. As the cortex continues to develop, secondary undulations,
which are shallower and more variable than the structures obtained in earlier
gyrification stage emerge, reproducing well-known characteristics of higher-order folding in the
mammalian, and particularly the human, brain.
\end{abstract}


\end{frontmatter}

\newcommand{\F}[0]{\textbf{F}}
\newcommand{\Pio}[0]{\textbf{P}}
\newcommand{\Int}{\int\limits}
\newcommand{\avg}[1]{\left<#1\right>}


\newcommand{\comment}[1]

\newcommand{\Jens}[1]{}
\newcommand{\Svenja}[1]{}
\newcommand{\Lucas}[1]{}

\section{Introduction}
\label{introduction}

One of the most striking features of the human brain is its highly folded
structure. Indeed, neuroscientists have for a long time pondered about its
importance and origin~\citep{bischoff_grosshirnwindungen_1868,
cunningham_complete_1890}. However, the process by which folds form is not yet
fully understood, neither as a mechanical~\citep{Bayly2014} nor as a molecular
process~\citep{Sun2014}. One of the main hypotheses to explain the convoluted
nature of the cortex, commonly called the \emph{differential tangential growth
hypothesis}~\citep{Richman} posits that the buckling of the brain is created by
a mismatch of growth rates in the cortical plate and the white matter
substrate. The main contention with this hypothesis, however, is its
requirement of a large difference between the stiffness of the two
regions~\citep{Bayly2014}. In order to obtain the wavelengths compatible with
the gyral width of the human brain, the initial form of the differential
growth hypotheses requires the stiffness ratio between the gray and white
matter to be in the order of 10~\citep{Richman}. This is a major hurdle, as
currently there is no consensus if the gray matter is
indeed stiffer than the white matter, and if so, by how much. There is a solid
body of evidence supporting the two possibilities, i.e., that the gray matter
is indeed stiffer than the white matter~\citep{Budday2015a, Green2008,
Johnson2013}, and vice-versa~\citep{Manduca2001, McCracken2005,
vanDommelen2010}. A second issue with the differential growth hypotheses is the
shape of the sulci. This model results in smooth sinusoidal patterns,
while the brain is characterized by smooth gyri and cusped
sulcii~\citep{tallinen_gyrification_2014}.

The study of layered systems has been extensively conducted in the field of
engineering, where it was used to model the buckling of sandwich-type
panels~\citep{hoff_buckling_1945}, the Earth's crust~\citep{biot_folding_1957,
ramberg_folding_1970}, etc. These works, however, deal mostly with stiff
materials and large stiffness ratios.  In recent years, there have been a surge
in the number of works dealing with soft materials, with special focus in
bio-compatible applications~\citep{budday_wrinkling_2017,
vandeparre_hierarchical_2010} which provides an important tool in the
understanding of the role of mechanics in the folding of the mammalian cortex.

Much work has been done to solve the issues with differential growth, specially
on the sulci formation. For instance,
\cite{tallinen_gyrification_2014, Tallinen2016} performed large simulations
to understand how the geometry and constraints affect the cortical folding,
where they showed that the size and shape of the folds are dictated by the
geometry of the early fetal brain. Other hypotheses have also been proposed to
explain cortical gyrification. Perhaps best known in the medical
community,~\cite{Essen1997} conjectured that axonal traction is the driver of
folding. Reaction-diffusion models, where the concentration and diffusion of
growth-activator chemicals are explicitly modeled, have also been suggested as
a way to explain both the gyrogenesis process, as well as the growth profile
itself~\citep{hinz_isogeometric_2019, Verner}.

Despite all these efforts, an aspect of brain folding that still remains
elusive is the phenomenon of hierarchical folding, an important feature of the
brain development, which must be included in order to understand the driving
forces behind the complex folding patterns observed in the human brain.

Recent studies have analyzed the influence of growth and stiffness
inhomogeneities along the cortex.  A few years ago, Toro et al. performed studies on the
effect of cortical inhomogeneity and curvature~\citep{Toro2005}. More recently,
Budday et al.~\citep{Budday2015193, Budday2018} performed similar inhomogeneity
studies on rectangular geometries. In these works, structures with resemblance to
the higher-order folding were obtained, but they lacked the complex spectrum of
folding present in the human brain.

It has been shown that the thickness of cortex of the brain impacts the width and
structure of brain folds~\citep{armstrong_ontogeny_1995, budday_role_2014,
heuer_role_2019, Mota74, Richman, Toro2005}. However, evidence on how
competition between the different thicknesses in the cortex affect folding has
been lacking. In this paper, the differential tangential growth
hypothesis is augmented with an inhomogeneous cortical thickness field,
yielding realistic folded structures, which could help explain formation of the
deep sulci in the mammalian brain, hierarchical folding, as well as its
consistent localization.


\section{Material and Methods}
\label{materials-and-methods}

\begin{figure}[!htb]
\centering
\includegraphics[width=80mm]{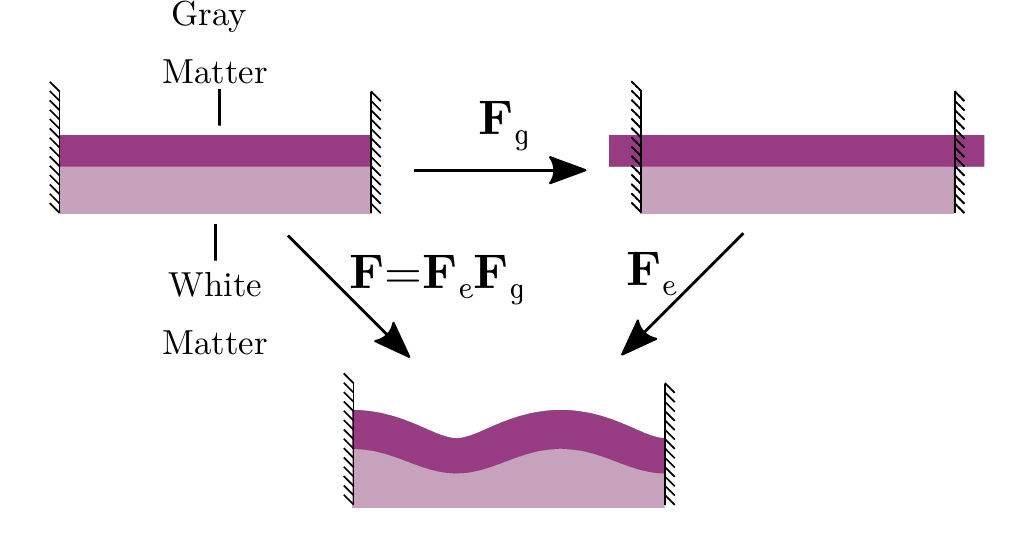}
\caption{(Color online) Schematic representation of the model. The
purple layer atop mimics the gray matter and is grown tangentially, while
the pink substrate underneath mimics the white matter and does not grow.
Growth is mathematically represented by the growth tensor $\F_g$, which can be
discontinuous.
In order to keep the compatibility with the attachment constraints between the
gray and white matter, the system is subject to residual stress, described in
this framework by the $\F_e$ tensor. \label{fig:model_cte}}
\end{figure}

We analyze two-dimension systems composed of two layers: a
purely elastic, non-growing, softer substrate in the lower region, mimicking
white matter, and a growing stiffer region on the top, emulating the cortical
gray matter (see Fig.~\ref{fig:model_cte}).
The simulations are performed
using a custom written finite element Method code to solve the continuum
mechanics equations\footnote{The simulations were written using the deal.II
library~\citep{dealII90, dealII10}, and parallelized using MPI via
PETSc~\citep{petsc-web-page, petsc-user-ref}.}.

\subsection{Theoretical background}
\label{sub:theorerical_bg}

Due to the large-strain, nonlinear nature of the human brain, the framework of
continuum mechanics~\citep{wood} is used. In order to distinguish between the
original and deformed configurations, the following notation is introduced: The
vector $\textbf X$ denotes the coordinates of the original configuration,
$\textbf x$ the coordinates of the deformed configuration, and $\textbf{u} =
\textbf{x} - \textbf{X}$ denotes the displacement field. The deformation
gradient tensor is written as

\begin{equation*}
\F = \frac{\partial\textbf{u}}{\partial\textbf{X}} + \textbf{I},
\end{equation*}
where $\textbf{I}$ is the identity matrix.
Growth is introduced using Rodriguez theoretical framework~\citep{rodriguez},
where the deformation gradient tensor $\F$ is decomposed into $\F = \F_e \F_g$
(see Fig.~\ref{fig:model_cte}), where , $\F_e$ describes the
elastic part of the deformation and $\F_g$ the growth contribution.
The energy and stress are calculated from the elastic part of the deformation
gradient tensor alone. Thus, in this framework, the energy-density (and all
quantities derived) is defined in terms of $\F_e$.

The soft tissue of the brain is modeled by the compressible Neo-Hookean energy-density function~\citep{wood}

\begin{equation}
\psi(\F_e) = \frac{\mu}{2} \left(\text{tr}(\F_e^T\F_e)- 2\log(J_e) - 2\right) + \frac{\lambda}{2} \log^2(J_e)\label{eq:neohookean}
\end{equation}
where $J_e = \det (\F_e)$, and $\mu$ and $\lambda$ are the Lamé parameters.
This energy-density family has been shown to appropriately model the brain
tissue~\citep{Budday2015193, budday_mechanical_2017}.

Due to the relatively long time scale of cortical development when compared to
the elastic response of brain tissue, the quasi-static
approximation is used. At every value of $\F_g$ the displacement
field $\textbf{u}$ is calculated, obeying the equilibrium equation

\begin{equation}
    \nabla \cdot \textbf{P} = \textbf{0},
    \label{eq:equilibrium}
\end{equation}
where $\textbf{P}$ is the first Piola-Kichhoff stress tensor, related to the
energy-density $\psi$ by

\begin{equation}
     \textbf{P} = \frac{\partial \psi}{\partial \F_e}.
     \label{eq:piola}
\end{equation}

Both, Eq.~\ref{eq:equilibrium} and Eq.~\ref{eq:piola} retrieve the functional
form of their classical continuum mechanics counterparts in the limits of no
growth, i.e., $\F_g = \textbf{I}$.
The values of $\lambda$ and $\mu$ are chosen such that in the linear (i.e.,
small strain) regime, the Young moduli ratio between the gray matter (GM) and
the white matter (WM) is $E_{\mathit{GM}}/E_{\mathit{WM}} = 3$, consistent with previous
models~\citep{Budday2015193} and the Poisson ratio $\nu = 0.35$ on both layers.
Due to the nonlinear nature of Eq.~\ref{eq:neohookean}, the value of the
Poisson ratio $\nu$ and Young modulus are dependent on the current displacement
in the system. Specifically, they depend on the determinant $J_e$ \citep{wood}
as

\begin{equation*}
\nu = \frac{1}{2}\frac{\lambda}{\lambda [1- \log(J_e)] +
\mu},\label{eq:poisson}
\end{equation*}
and
\begin{equation*}
    E = \frac{\mu - 3\lambda \log(J_e)}{J_e} \frac{2\mu + \lambda(3-2\log(J_e))}{2\mu
    + \lambda(1-\log(J_e))}.
\end{equation*}

Notably, the energy-density in Eq.~\ref{eq:neohookean} has no inherent length scale.
Thus, only the ratios between the elastic moduli are important for the
phenomenology presented in this paper.


\subsection{Simulation details}

The cortical layer is grown linearly, i.e., the growth tensor is described by

\begin{equation*}
    \F_g(\theta_g) = \theta_g \textbf{I} + (1-\theta_g) \hat X_y \otimes \hat X_y\label{eq:growth}
\end{equation*}
where the growth parameter $\theta_g$ measures the degree of elongation in the
cortex. For instance, at $\theta_g =2$ the gray matter would have expanded to
twice its lateral size if it were not constrained. To mimic differential
growth, $\theta_g$ is varied in the interval $[1.0, 2.5]$ in the gray matter
region, while it is kept at unity in the white matter region. The growth
parameter $\theta_g$ is increased in small steps of 0.01.

At every growth step, Eq.~\ref{eq:equilibrium} is solved using the finite
element method, with boundary conditions of zero displacement on the bottom surface
$X_y = 0$ and zero stress on the top surface $X_y = L_b$. In order to
minimize boundary effects, periodic boundary conditions are imposed on the
sides of the surface, $X_x = 0$ and $X_x = L_b$. The box lengths $L_b$ will
be specified in each section.
The system being two dimensional, corresponds to an infinite system in the
$z$-axis, with the constraint of no displacements in the $z$-axis.

Due to the nonlinear nature of the energy described in Eq.~\ref{eq:neohookean},
the divergence of the first Piola-Kichhoff stress tensor will also be
nonlinear. To find the roots of this function, the Newton method augmented by
a backtracking algorithm~\citep{wood, Nocedal} is used. In order to avoid
overlaps, collisions are detected and resolved using the approach introduced
by~\cite{mcadams_efficient_2011}. As any collisions will be initiated in the
cortical region, calculations are optimized by only detecting collisions in the
gray area elements. This generates no artifacts, as due to the structure of the
mesh, collisions are initiated in the gray area, and are resolved before any
white matter elements are involved.

In order to allow the system to overcome metastable states,
a small force field pointing in the $X_y$ direction is introduced. The
forces are drawn from a uniform random distribution between
[$-7\times10^{-2}, 7\times 10^{-2}$)$\times E_{\mathit{GM}}$. Each
simulation has been repeated three times with different seeds.
No significant difference is observed between runs with different seeds, or
when the forces are doubled or halved.


\section{Results}
\label{results}

\subsection{Homogeneous Thickness of Cortical Ribbon}
\label{sec:constant}

\begin{figure}[htbp]
\centering
\includegraphics[width=80mm]{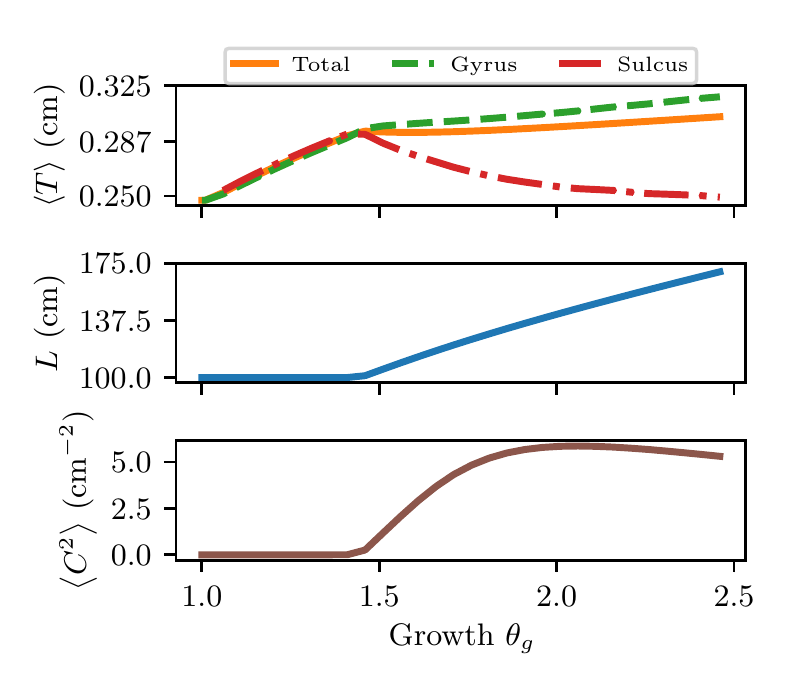}
\caption{(Color online) Spatial averages of several observables in the system
with initial thickness $T$ = 0.250 cm, as a function of the growth parameter
$\theta_g$. In (a),
the dependence of the average thickness with growth are shown, as well as the
average cortical thickness of the sulci and the gyri, as defined in the text.
In (b) and (c) the contour length of the top layer of the system and
squared curvature of the system are shown, respectively. This last quantity is specially
useful to characterize the onset of folding, as will become clear in Sec.
\ref{sec:inhomogeneous}.
\label{fig:quant_homo}}
\end{figure}

We analyze the folding of a slab with constant cortical thickness $T$
throughout, in the range $[0.1, 0.5]$cm. In this case, it is expected that the
system will fold into well defined wavelengths, with no
localization~\citep{groenewold_wrinkling_2001}. In order to avoid finite-size
effects, the simulation box is set to $L_b=100 \text{cm} \gg \lambda_F$, where
$\lambda_F$ is the folding wavelength.

\begin{figure}[!htb]
\centering
\includegraphics[width=80mm]{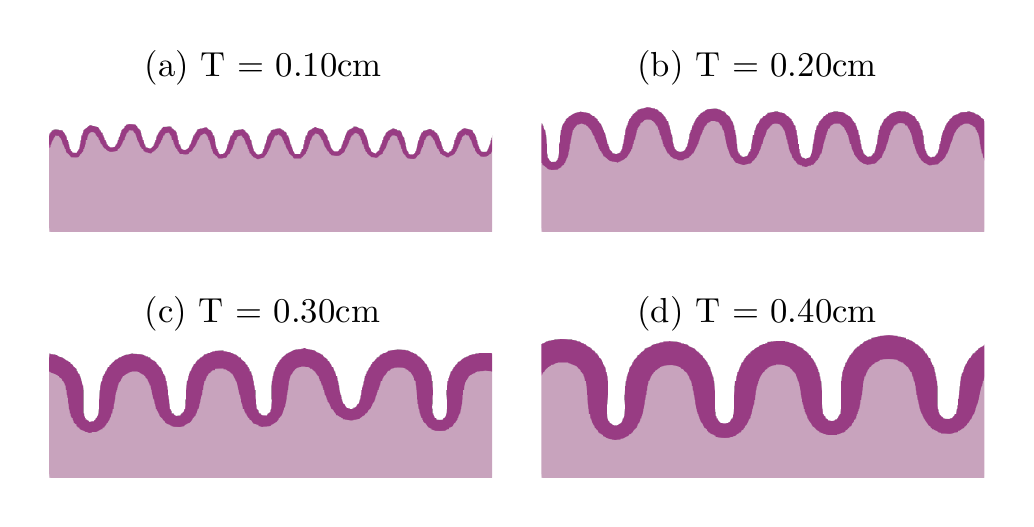}
\caption{(Color online) Folded system for the thickness indicated within
the figure. In order to improve visualization, only the region
$y > 95$cm is shown here.\label{fig:folds_cte}}
\end{figure}

Initially (i.e., for little growth) even though the growth happens
tangentially, the cortex does not increase in length, but rather in thickness
due to the confinement and resulting stress (see Fig.~\ref{fig:quant_homo}).
Eventually, at a critical growth $\theta_g^C$, the stress exceeds the critical
buckling threshold, and the system buckles in a well defined, almost sinusoidal
wave pattern (see Fig.~\ref{fig:folds_cte}). As growth continues, the
wavelength of the pattern and the average thickness remain almost unchanged,
while the amplitude increases. Interestingly, the thickness of the cortex is
no longer homogeneous. Sulci (regions of positive curvature) are significantly
thinner than gyri (regions of negative curvature). As growth continues post
buckling, the difference in thickness continues to increase.

In each simulation the Fourier
transform of the function $u_y(X_x)$ (i.e., the displacement in the $y$
direction as a function of the material coordinates) is calculated along the
top layer of the system. The weighted average wavenumber
and wavelength are then obtained as

\begin{equation*}
    \avg k = \frac{\sum_k k |h^2_k|}{\sum_k |h^2_k|} \quad \avg{\lambda_F}= \frac{2
    \pi}{\avg k},
\end{equation*}
where $h_k$ is the coefficient of Fourier expansion for the mode with
wavenumber $k$.
In order to obtain the weighted average wavelength for each cortical thickness $T$, the
simulations are repeated with inceasing number of mesh cells, ranging from $2^{12}$
to $2^{20}$ cells. The value of the weighted average wavelength is then obtained by
wavelength via a linear extrapolation to the infinitely refined mesh.

\begin{figure}[htbp]
\centering
\includegraphics[width=80mm]{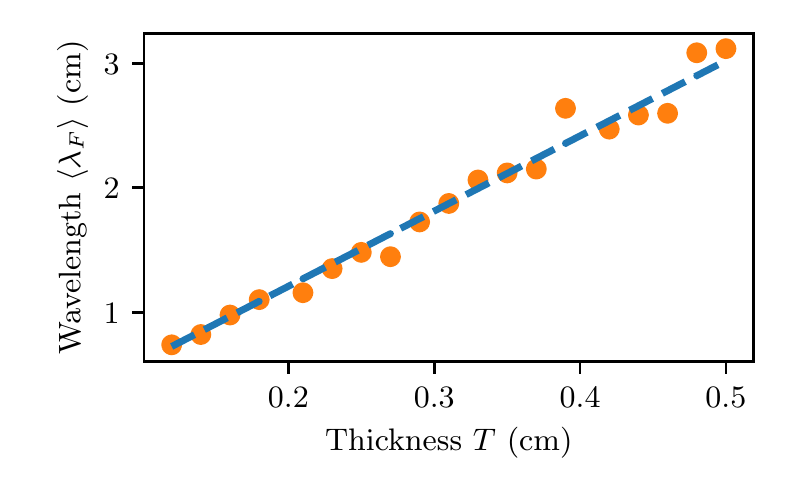}
\caption{(Colors Online) Weighted wavelength of the system as function of the
cortical thickness $T$. The orange circles indicate extrapolation results,
while the broken blue lines represent the linear fit $\avg{\lambda_F} =
6.04 T$ cm with $R^2=0.98$. \label{fig:wavelength}}
\end{figure}

\begin{figure}[!htb]
\centering
\includegraphics[width=80mm]{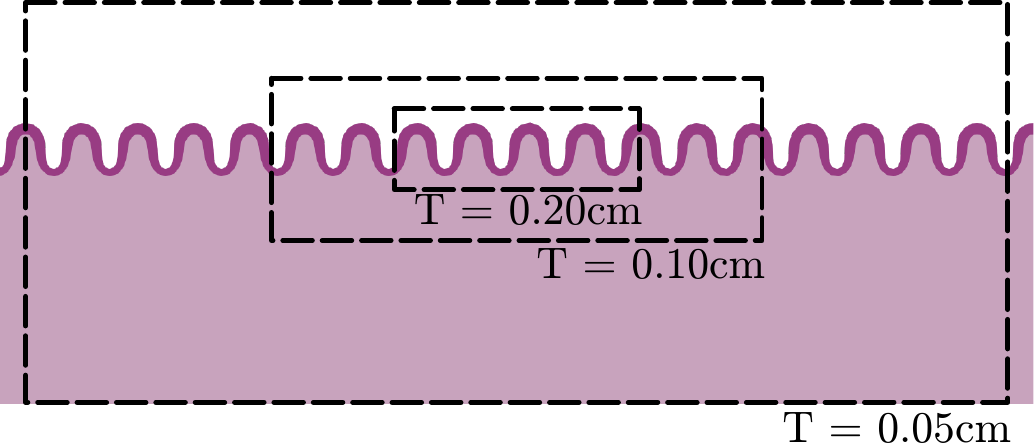}
\caption{(Color online) Schematic representation of scaling of the system.
Due to the lack of inherent length scale in the elastic equations, systems
with thicker cortices can be seen as subregions of systems with thinner
cortices. Here, each broken rectangle highlights a region which is
equivalent to a system with the cortical thickness $T$ indicated within the
figure.
\label{fig:zoom}}
\end{figure}

The weighted average wavelength increases linearly with initial thickness $T$
(see Fig.~\ref{fig:wavelength}). The linear dependency can be obtained from
simple dimensional analysis: As the elastic equations have no inherent length
scale, the cortical thickness is the only relevant length scale, as long as
system size is not limiting. Thus, changing the homogeneous cortical
thickness can be seen as a change of measurement units, or progressive zooming
in on the same base system (see Fig.~\ref{fig:zoom}). It is possible to
estimate the slope for a linear elastic substrate through analytical
calculations as~\citep{groenewold_wrinkling_2001}

\begin{equation}
    \lambda_F(T) = \pi T \left(\frac{2}{1-\nu^2}
    \frac{E_{\mathit{GM}}}{E_{\mathit{WM}}}\right)^{1/3} \approx 5.96 T.
    \label{eq:analytical_wavelength}
\end{equation}

Such slope presents a weak dependence on the stiffness ratios between the gray
and white matter, obeying a weak power law. Thus, the specific value of the
ratio $E_{\mathit{GM}}/E_{\mathit{WM}}$ plays only a minor role in the
determination of the folding wavelength.

\subsection{Inhomogeneous Thickness of Cortical Ribbon}
\label{sec:inhomogeneous}


The cortical thickness of the brain is spatially inhomogeneous. In order to
emulate this inhomogeneity, a variable cortical thickness $T(X_x)$ is
introduced. Specifically, as a generic form of thickness variation, a
sinusoidal thickness variation of the form
\begin{equation*}
  T(X_x) = A\sin(2\pi X_x/L_t) + T_0
\end{equation*}
is chosen. This inhomogeneity introduces two new length scales beyond the
base thickness $T_0$: the inhomogeneity amplitude $A$ and the period of the
inhomogeneity $L_t$.
Thus, in contrast to the previous results, it is possible to choose any one of the three
as the fixed length scale, and vary the remaining two independently. In this
study, the thickness period $L_t=10$ cm is chosen as the fixed scale. The
folding pattern for a different periodicity $L_t'$ can be obtaining by
rescaling the spatial quantities by $L_t'/L_t$, or a suitable power thereof.

Note that any form of thickness variation can be written as a sum of sinusoidal
variations. When deformations are small, even the resulting folding patterns
can be obtained by simple superposition. In the brain, however, deformations are
large and nonlinear, and each thickness field must be studied independently.

Simulations are performed for base thickness in the same range as before, [0.1,
0.5] cm, and for each $T_0$, the amplitude $A$ was varied in the range
$[0, 0.9] \times T_0$. The inhomogeneity creates a much more localized
deformation, thereby reducing possible artifacts created by finite-size
effects. Thus, in order to maximize computational efficiency, the simulation
box is chosen as $L_b = L_t$.

\begin{figure}[!htb]
\centering
\includegraphics[width=80mm]{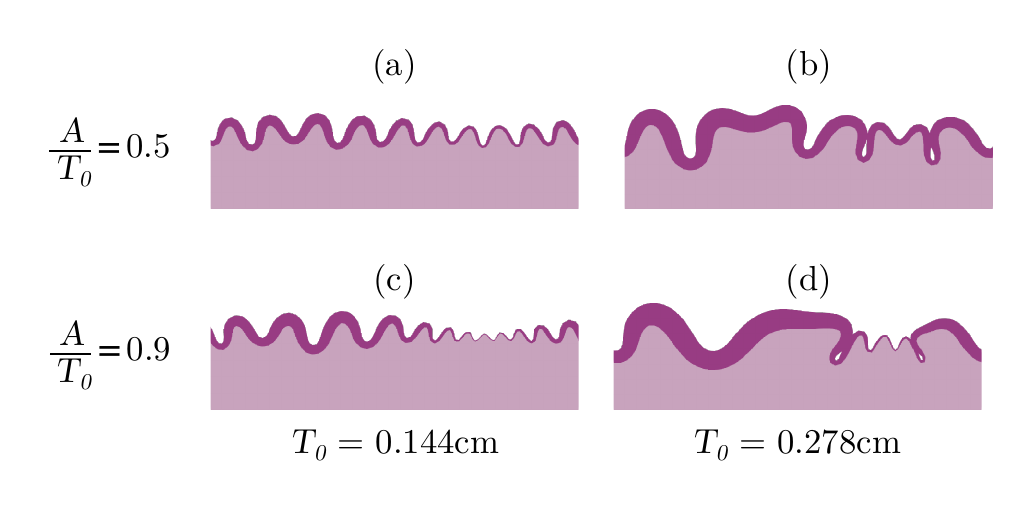}
\caption{(Color online) Simulations with varying inhomogeneities
amplitudes $A$ at growth parameter $\theta_g = 2.5$. The simulations have a
base thickness $T_0 $ as indicated within the figure.\label{fig:2D_amplitudes}}
\end{figure}

If the natural folding wavelength $\lambda_F$ of the local thickness is much
smaller than the periodicity length, the system behaves essentially like the
homogeneous systems  studied in Section~\ref{sec:constant}, i.e., the folding
wavelength obeys Eq.~\ref{eq:analytical_wavelength}, with $T = T(X_x)$. Here, the system
folds into well defined waves, but with the spatially dependent wavelength
commensurate with the cortical thickness of the underlying region (see
Fig.~\ref{fig:2D_amplitudes} (a), (c)). Accordingly, these systems also present
the constant length - constant thickness stress-relief mechanism.

However, when the folding wavelength becomes comparable with the periodicity
length, a second form of folding arises, characterized by complex folding
patterns. In these conformations, several wavemodes are simultaneously obtained
(see Fig.~\ref{fig:2D_amplitudes} (b), (d)), presenting similarities with the
further regions of the gyrencephalic brain.

\begin{figure}[tb]
\centering
\includegraphics[width=80mm]{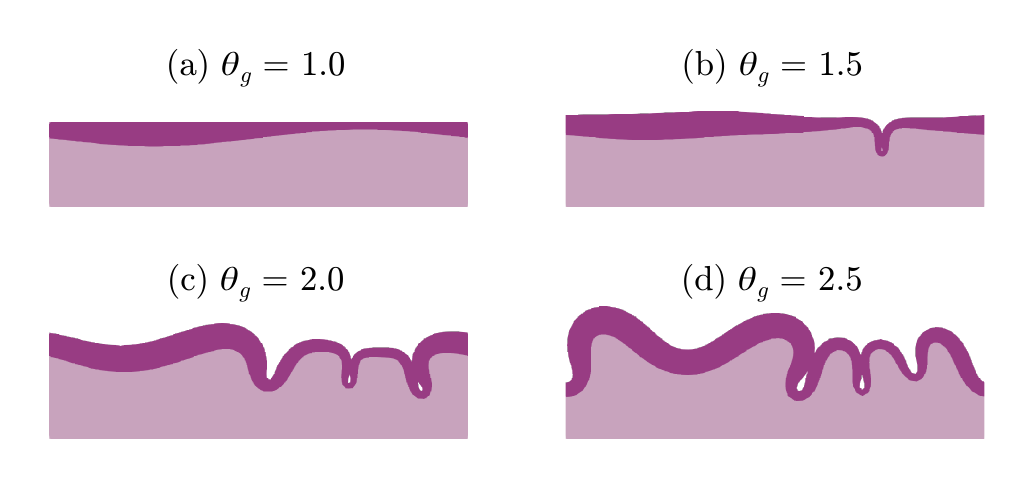}
\caption{(Color online) Growth evolution of system with
$T_0 = 0.45$cm and $A = 0.315$ cm. The growth parameters are indicated within the
figure.\label{fig:2D_time}}
\end{figure}

This new shape has distinct developmental steps, which differ from those
described in Sec.~\ref{sec:constant}.  For small growth, a initially flat
system (see Fig.~\ref{fig:2D_time} (a)) forms a single, deep, sulcus in an
otherwise planar cortex, in the region surrounding the thickness minimum (see
Fig.~\ref{fig:2D_time} (b)).  The depth of this sulcus soon saturates, and due
to the underlying white matter, it is energetically favorable to form
additional sulci rather than to increase the depth of the exist sulcus as
growth continues (see Fig.~\ref{fig:2D_time} (c)). The maturation of the new
sulci occur concurrently with the formation of shallow sulci in the regions of
highest thickness.(see Fig.~\ref{fig:2D_time} (d))

It is expected that the folding starts on the region of thinnest cortex. In the
limit of small deformations, the system can be analysed by the theory of thin
plates. In this domain, the bending rigidity depends on the cube power of the
thickness of the plate~\citep{ventsel2001thin}. Thus, the large differences in
thickness create a stress imbalance in the region, leading to the buckling of
the region with small thickness. This is specially noticeable in the formation
of the deep sulci observed in Fig.~\ref{fig:2D_time}. Here, the thick parts
of the cortex compress laterally, which leads to stress condensation in the
thin parts of the cortex. The thin region has then to absorb the compression of
the whole system.

The simulations are in qualitative agreement with the results from linear
stability analysis. The details how this theory is applied to our model are
outlined in~\ref{sec:analytical_model}. In short, the lowest-energy mode of a
bilayer system is calculated, where we introduce a thin plate as the top
layer, and an elastic substrate underneath.  The upper plate has a sinusoidally
varying bending rigidity
\begin{equation*}
    \kappa(x) = \kappa_0 + \kappa_1 \sin\left( \frac{2\pi}{L} x\right),
\end{equation*}
where $L$ is the periodicity length. The system is then subject to a spatially
constant compressive surface tension $\gamma$. The bending rigidity plays a
similar role in this model as the thickness plays in the simulations, and the
surface tension plays a similar role to growth.

\begin{figure}
\centering
\includegraphics[width=80mm]{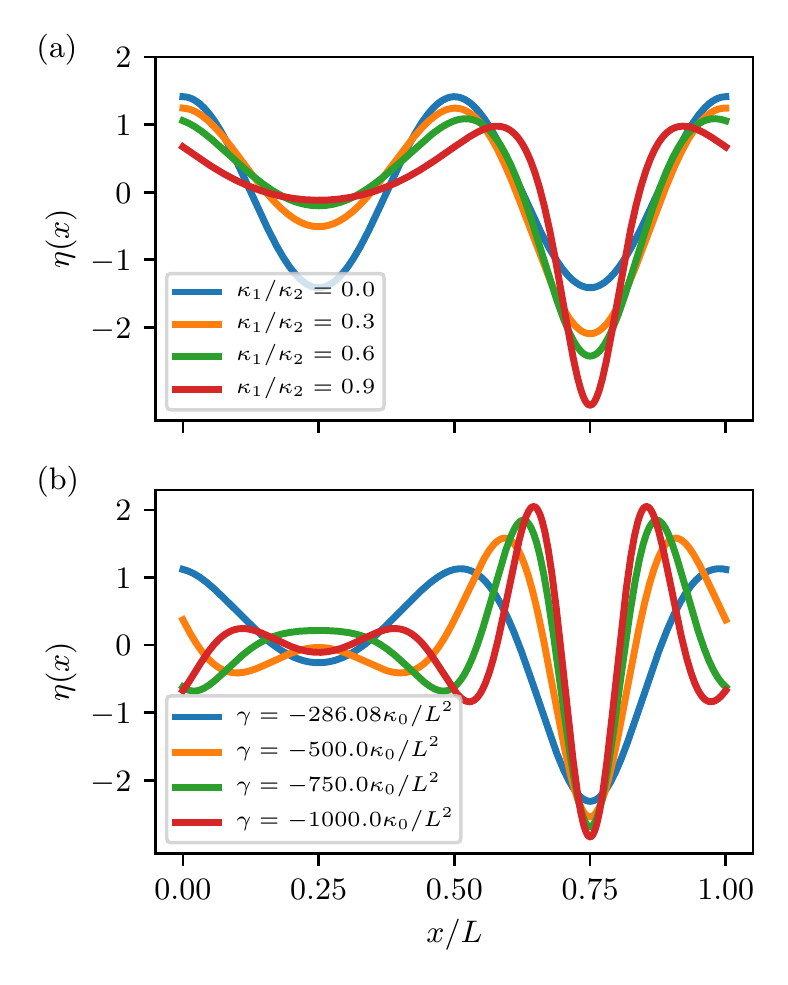}
\caption{(Color online) Analytical prediction of vertical dislocations $\eta$.
(a) Varying bending rigidity ratios at buckling point. (b) Varying values
of surface tension $\gamma$, while the bending rigidity ratio is kept
constant as $\kappa_1/\kappa_0 = 0.5$. For $\gamma > -286.08\kappa_0/L$, no
buckling is predicted. Note the appearance of higher order folding for
$\gamma < -1000 \kappa/L$. In both cases, the effective Young Modulus used
was $\hat E = 2000 \kappa_0/L^3$. As the curves are normalized, the
predicted dislocations can only be compared within the same curve, but not
between curves calculated with distinct parameters.
\label{fig:buckling_theory}
}
\end{figure}

According to the analytical model, homogeneous systems (i.e., $\kappa_1 = 0$)
will fold into a single, well-defined wavenumber, as expected (see
Fig.~\ref{fig:buckling_theory} (a))~\citep{hornung_simulation_2019}. However,
as the bending rigidity ratio $\kappa_1/\kappa_0$ is increased, the folding
gets more localized around the bending rigidity minimum. This phenomenon is
qualitatively consistent with what was observed in the simulations. For the
case with $\kappa_1/\kappa_0 = 0.9$, for instance, the linear stability
analysis predicts the formation of a deep sulcus surrounding the bending
rigidity minimum, similarly to the finite-element simulations (see
Fig.~\ref{fig:2D_time} (b)). The analytical model predicts that as the cortical
plate gets more compressed, the system develops secondary sulci, as can be
noticed in region $x/L \approx 0.25$ in the system with $\gamma = -1000
\kappa_0/L$ (see Fig.~\ref{fig:buckling_theory} (b))

\begin{figure}
\centering
\includegraphics[width=80mm]{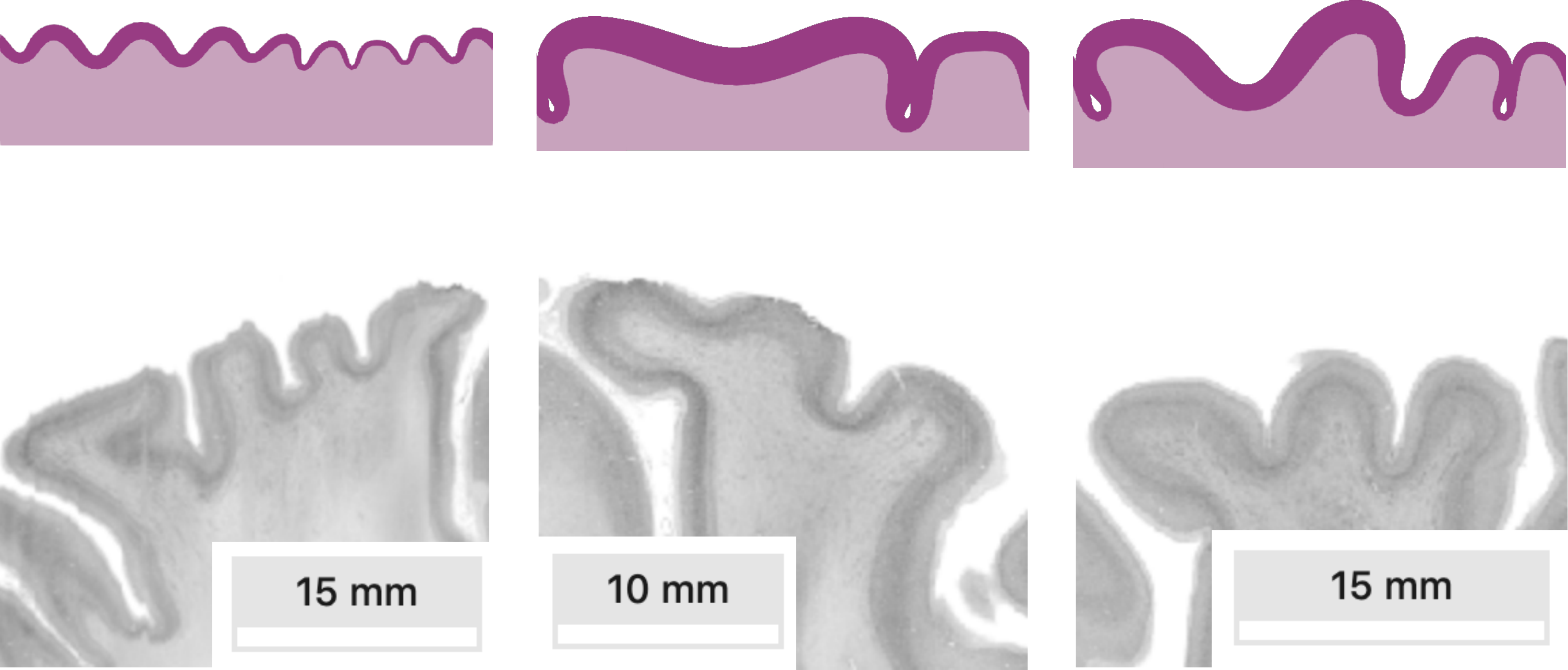}
\caption{Illustrative comparison between simulation results (top) and sections
of the cortex (bottom; adapted from HBP BigBrain~\citep{bigbrain}). From
left to right, the simulations are performed with $T_0 = 0.189$ cm,
$A=0.102$ cm; $T_0 = 0.500$ cm, $A=0.180$ cm; $T_0 = 0.367$ cm, $A=0.198$
cm. In the same order, extracts of the left superior parietal lobule
(sagittal plane), the right postcentral gyrus (coronal), and the right
posterior middle temporal gyrus (coronal) are shown. Due to the arbitrary
choice in the value of $L_t$, the thicknesses between the simulation and the
histological section are not quantitatively comparable.
\label{fig:biocomparison}}
\end{figure}

It is possible to compare the structures obtained to histological sections of
the human brain, as shown in Fig.~\ref{fig:biocomparison}. The results from our
simulations present a striking similarity to some regions of the human cortex,
showing a wide range of sulcal depths and widths, in qualitative agreement with
the ones observed in regions with higher-order folding.  For instance, the
superior parietal lobe presents a plethora of small, shallow folds, similar to
those observed in the simulations with thin cortices. Regions presenting a more
complex folding pattern, such as the postcentral gyrus, or the posterior middle
temporal gyrus are reproduced by simulations with thicker cortices.
Furthermore, the gyral height-to-width ratio resulting from the simulations are
similar to those observed in the histological sections.

\begin{figure}[tb]
\centering
\includegraphics[width=80mm]{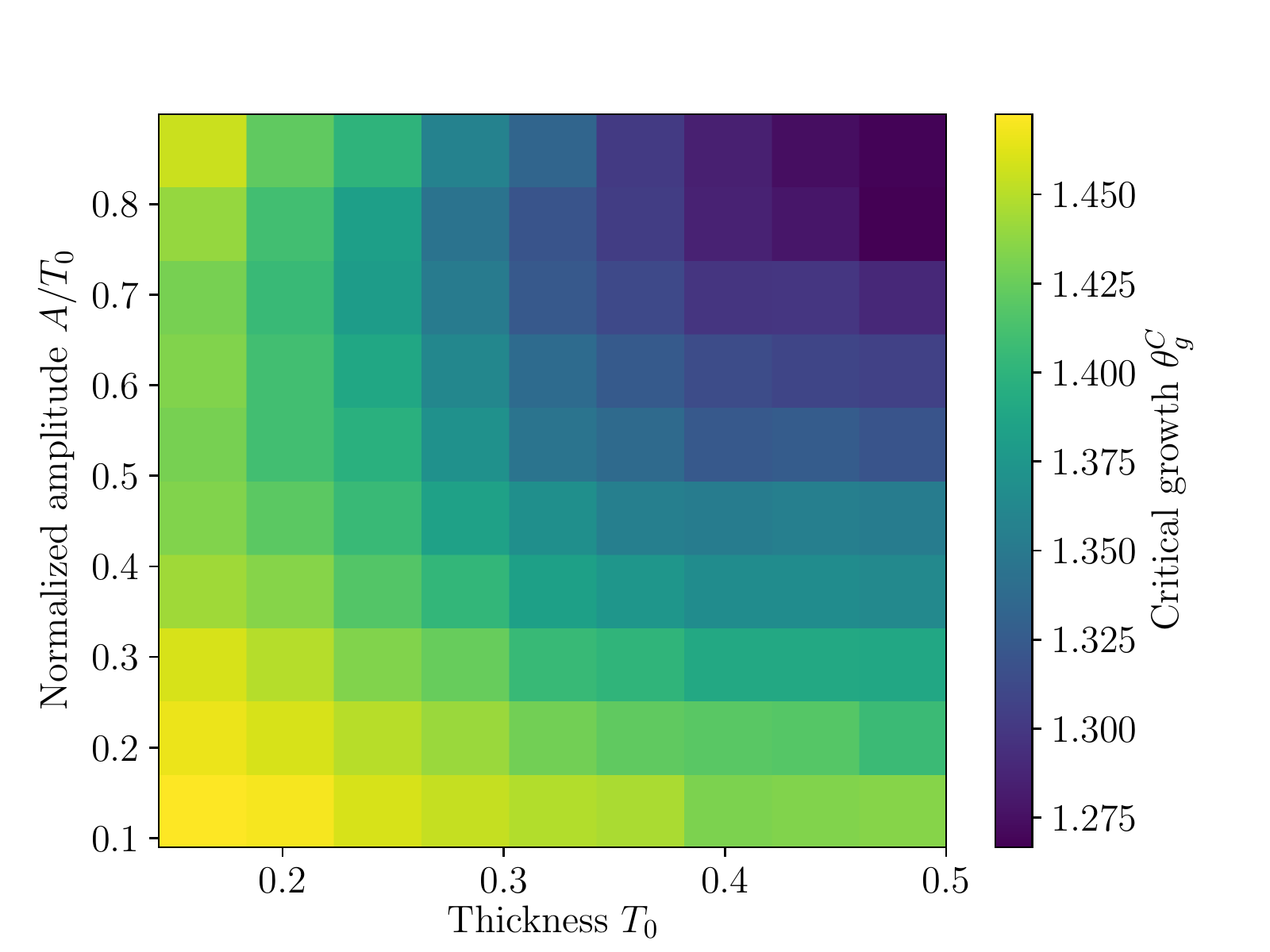}
\caption{(Color online) Critical growth $\theta_g^C$ for the emergence of
folding for various combinations of cortical thickness and inhomogeneity
amplitudes. \label{fig:2D_phasespace}}
\end{figure}

Next, we turn our attention to the onset of buckling, i.e. the critical amount
of growth $\theta_g^C$ above which the system starts to fold, broadly
indicating when the
constant-length regime ceases, and the constant-thickness starts. The critical
growth $\theta_g^C$ is defined somewhat arbitrarily as the growth where the
averaged squared curvature of the system reaches 2cm$^{-2}$. The results do not
change considerably for choices of critical curvature square $\left<C^2\right>$
in the range [1, 3]cm$^{-2}$.
The transition points are shown in Fig.~\ref{fig:2D_phasespace}, where it is
possible to observe that the critical growth $\theta_g^C$ is strongly affected
by interplay between the inhomogeneity amplitude and the base thickness of the
cortical plate, with a noted decrease in the value of $\theta_g^C$ as the
inhomogeneity amplitude $A$ increases. In light of these results and of the
differences between the folding of thin and thick systems (see
Fig.~\ref{fig:2D_amplitudes}), we conjecture that the relevant parameter to
cortical convolution is not solely the ratio between the maximal and minimal
thickness, but that the local gradient of the cortical thickness also plays a
fundamental role.

\section{Conclusion}
\label{concluding-remarks}

We have analyzed the effects of the cortical inhomogeneity in the formation of
brain folds. To this end, two closely related systems were studied. First,
analyses was carried out by simulating a rectangular bilayer slab where the top
layer grows tangentially. It was observed that the folding pattern follows
well-defined wavelengths, which depended on the thickness of the top layer,
consistent with previous work~\citep{biot_bending_1937, budday_role_2014,
groenewold_wrinkling_2001}.
According to Bok's principle, the thickness differences between the sulci and
gyri are created as a consequence of the curvature of the brain. During
gyrogenesis, the high curvature in the sulci spreads the cortical mantle,
decreasing its thickness. The gyral crowns are not affected as strongly due to
their relatively small curvature~\citep{bok_einflus_1929}. Our simulations are
consistent with this principle, showing that an initially homogeneous cortex
can develop cortical inhomogeneities through buckling, coherent with prior
observations on homogeneous systems~\citep{holland_symmetry_2018,
riccobelli_surface_2020, welker_why_1990}.

Second, the effects of wavelength competition were studied through the
introduction of inhomogeneities in the cortical thickness. In these systems,
phenomena closely related to the mammalian gyrogenesis, such as the emergence
of hierarchical folding in systems with thick cortices, were observed. Indeed,
the results shown here indicate that inhomogeneities in the cortical thickness
might play an important role in the localization and formation of hierarchical
folding patterns of the brain. Specifically, it was shown that these
inhomogeneities are sufficient to break the simple wave-like patterns observed
in the homogeneous system.  Further, our observations indicate that thickness
inhomogeneity leads to earlier folding compared to systems with homogeneous
cortical thickness.  Lastly, the results obtained \emph{in silico} were proved
to be consistent with those obtained from analytical models derived from thin
plate theory.  Similar approaches have been taken in other studies, where other
cortical inhomogeneities were studied both in circular~\citep{Toro2005} and
rectangular~\citep{Budday2018} geometries. Our results are in line with those
findings, but exhibiting a more complex folding pattern, as well as the
emergence of multiple sulci within each inhomogeneity period.

The computational model used to explain the hierarchical folding patterns of
the mammalian brain is very general, utilizing only the fundamental elastic
nature of brain tissue. The absolute values of the elastic moduli play no role,
with only their ratios being important. This model is agnostic to all sorts of
structural properties of the brain, such as its volume, eccentricity,
functional connectivity, etc. The results derived here are thus applicable to
the brains of other mammalians, after a suitable scaling of the thickness,
growth, and periodicity lengths.

While the comparisons presented here focus mostly on smaller regions of the
human brain, we conjecture that more extreme forms of inhomogeneity can lead to
the formation of the deep sulci obtained in the mature human brain. Larger
simulations are required to gauge the influence of thickness inhomogeneity on
the full brain. Based on our current results, the mutual influence of the
different kinds of inhomogeneity (e.g., thickness and growth) can also be
studied.

We have shown the consequences of cortical thickness inhomogeneities to brain
folding. What drives the development of these inhomogeneities is still a matter
of ongoing research. Features of a given brain area, such as its cortical
thickness are related to its function~\citep{geyer_microstructural_2013}.
Biomolecular and mechanical factors contribute to
that~\citep{kriegstein_patterns_2006, Sun2014}, and, based on our current approach,
could additionally be included in future models of brain folding.


It can also be shown that viscoelastic properties can affect the buckling
wavelength of bilayered systems ~\citep{biot_folding_1957}. Thus, it would be
interesting to further refine the current model with the viscoelastic
characteristics of growing tissues in general~\citep{ranft_fluidization_2010},
and of the brain in particular~\citep{budday_rheological_2017}. Finally,
structural connectivity, i.e., axons within the white-matter fiber tracts, have
been conjectured to be one of the drivers of folding~\citep{Essen1997}. How
they would influence the conformations found in this work could be investigated
further to understand the mutual influence of differential growth and axonal
tension in gyrogenesis.

This paper's focus was the developing brain, but in virtue of generality of the
model used, its results are extensible to other fields. For instance,
it has been shown that microscopic corrugated surfaces give rise
hyperhydrophobic surfaces, in the so called Lotus effect~\citep{gao_lotus_2006,
marmur_lotus_2004}. Our results can provide further insight into the self
assembly of these corrugations, and in the control of their properties.

Soft layered systems can be realized experimentally by gel slabs coated with
gels with different properties~\citep{auguste_role_2014,
budday_wrinkling_2017}. These systems have been used as simulacra for brain
folding~\citep{holland_symmetry_2018, Tallinen2016}, where they were able to
mimic the folding a 3D-printed human brain. Thus, the production of
samples with sinusoidal variation of the top-layer thickness would allow for
the experimental testing of our predictions.

\section{Acknowledgments}
\label{acknowledgements}

The authors gratefully acknowledge the computing time granted by the JARA-HPC
Vergabegremium on the supercomputer JURECA~\citep{jureca} at Forschungszentrum
J\"ulich. Simulations were additionally performed with computing resources
granted by RWTH Aachen University under project rwth0399. We would like to
thank Dr.~Claude J. Bajada for helpful discussions and feedback. This work
was further supported by a grant from the Initiative and Networking Fund of the
Helmholtz Association (SC) as well as the European Unions's Horizon 2020
Research and Innovation Program under Grant Agreement 785907 (Human Brain
Project SGA2; SC). LCC also acknowledges the support by the International
Helmholtz Research School of Biophysics and Soft Matter (IHRS BioSoft).

\appendix
\section{Simplified Analytical Model}
\label{sec:analytical_model}

In order to understand the buckling of the system, we use a simplified model
which can be solved analytically. Here, a thin plate with a spatially varying
bending rigidity is studied. This plate is attached to a linear elastic
substrate, filling the whole of the half-space $y<0$. In the limit of small
deflections and disregarding shearing, it is possible to write the displacement
of the system in the Monge representation as

\begin{equation*}
    \textbf u(x, 0) = (0, h(x)).
    \label{eq:monge}
\end{equation*}

Here, $h(x)$ indicates the local height of the plate along the $x$ axis. The
free energy of this system is composed of three terms:

\begin{equation}
    F = F_{bend} + F_{stretch} + F_{subs},
    \label{eq:energy_comp}
\end{equation}
where $F_{bend}$ is the free energy of bending the thin plate, $F_{stretch}$ is
the energy required in order to stretch the plate and $F_{subs}$ is
the energy of the deformed underlying substrate. Explicitly,

\begin{align*}
    F_{bend} &= \Int_0^L \kappa(x) \left(\nabla^2 h(x)\right)^2 dx \\
    F_{stretch} &= \Int_0^L |\nabla h(x)|^2 \gamma dx \\
    F_{subs} &= \frac{1}{2} \Int_0^{\infty}\Int_0^L \sum_{ij}\sigma_{ij} u_{ij}
    dxdy
    \label{eq:explicit_terms}
\end{align*}
where $\kappa(x)$ describes the space-dependent bending rigidity, $\gamma$ is
the surface tension on the superficial plate, $\sigma_{ij}$ are the
components of the Cauchy stress tensor, and $u_{ij}$ are the components of the
strain tensor $u_{ij} = \frac{1}{2} \left(\frac{\partial u_i}{\partial x_j} +
\frac{\partial u_j}{\partial x_i}\right)$. In order to keep the simplicity of
the model the bending rigidity of the plate, rather than its thickness, varies
sinusoidally.
That is,

\begin{equation*}
    \kappa(x) = \kappa_0 + \kappa_1 \sin\left( \frac{2\pi}{L} x\right) = \kappa_0 +
    \kappa_1 \sin(k^* x),
\end{equation*}
where $k^*$ is the characterstic wavenumber of the inhomogeneity.
Due to the periodic nature of our system, the stability analysis is easier to
carry out in Fourier space. Thus, the height function $h(x)$ is expanded into

\begin{align*}
    h(x) &= \frac{1}{L} \sum_k h_k \exp(ixk), k \in \frac{2\pi}{L} \mathbb{Z} \\
    h_k &= r_k \exp(i \phi_k)
\end{align*}
with $\phi_{k} = \phi_{-k}$. In this decomposition,
whole-plate dislocation(i.e., $h_0 = 0$) were disregarded.

To obtain the energy of the elastic substrate in the Fourier space, one has to
solve the problem of a linear elastic substrate with given surface deformation
in Fourier space, as derived in Ref.~\citep{groenewold_wrinkling_2001}. With
this solution, the free energy described in
Eq.~\ref{eq:energy_comp} is written as

\begin{equation}
    \begin{split}
    F = \sum_k &
            \left( \frac{1}{2} ( \hat E |k| + \gamma k^2 + \kappa_0 k^4) r^2_k
            \right.  + \\
    & \left. \frac{\kappa_1}{2} k^2 (k + k^*)^2  r_k r_{k+k^*}\sin(\phi_k - \phi_{k +
k^*}) \right)
    \end{split}
    \label{eq:energy_sin}
\end{equation}
with
\begin{equation*}
    \hat E = E \frac{ 2 \nu^2 - 10 \nu + 5}{2(4\nu - 3)^2 (\nu + 1)}.
\end{equation*}

We search for the buckling modes that are most unstable. That is, those with
wavenumber $k$ which minimize the energy in Eq.~\ref{eq:energy_sin}. Due to the
bound properties of the sinus, it is clear that the condition $\phi_k -
\phi_{k+k^*} = 3/2\pi + 2\pi n$ is necessary to obtain this minimum. Thus, the energy
is written as

\begin{equation}
    \begin{split}
        F = \sum_k &
              \left( \frac{1}{2} ( \hat E |k| + \gamma k^2 + \kappa_0 k^4) r^2_k \right.  \\
           -& \left. \frac{\kappa_1}{2} k^2 (k + k^*)^2  r_k r_{k+k^*}           \right) .
    \end{split}
    \label{eq:free_energy}
\end{equation}

Unstable modes can then be obtained by standard stability analyses.
Eq.~\ref{eq:free_energy} is recast into its matricial form

\begin{equation*}
    F = \frac{1}{2}\textbf r^T H \textbf r,
\end{equation*}
where $\textbf r$ is a vector with components $\textbf r = r_k$, and $H$ is the
Hessian matrix. Explicitly,

\begin{align*}
    H_{ij} =& 4 g(k_i) \delta_{k_i, k_j} - 2f(k_i) \delta_{k_j, k_i+ k^*}\\
    &- 2f(-k_i)(\delta_{k_j, k^* - k_i} - \delta_{k_j, k_i - k^*})
\end{align*}
with

\begin{align*}
    g(k) &= \frac{1}{2} ( \hat E |k| + \gamma k^2 + \kappa_0 k^4), \\
    f(k) &= \frac{\kappa_1}{2} k^2 (k + k^*)^2.
    \label{eq:polynomials}
\end{align*}
In this form, the unstable modes are obtained as those states with negative
eigenvalues for the Hessian matrix, corresponding to modes with negative
energy. For homogeneous systems (i.e., $\kappa_1=0$), the energy contribution
of each mode is independent (i.e., the Hessian matrix is diagonal), and modes
that minimize the energy can be obtained
analytically~\citep{hannezo_instabilities_2011, hornung_simulation_2019}. In
the inhomogeneous case, the various wavemodes are coupled, and it is necessary
to solve the eigenproblem numerically. Fig.~\ref{fig:buckling_theory} shows
the results of these calculations. Each curve corresponds to the eigenfunctions
with lowest eigenvalues for different elastic parameters, as indicated therein.

This analytical theory gives results which are qualitatively similar to those
obtained in our simulations. In order to obtain quantitative comparisons, a
more complex theory is necessary, which takes into account the lateral
displacements during gyrification.

\addcontentsline{toc}{section}{References}

\bibliography{higher_order_folding}{}
\bibliographystyle{elsarticle-num-names}

\end{document}